\documentclass[%
 reprint,
 amsmath,amssymb,
 aps,
]{revtex4-2}

\usepackage{graphicx}
\usepackage{dcolumn}
\usepackage{bm}
\usepackage{physics}
\usepackage{color}
\usepackage{url}

\usepackage{mleftright} 

\usepackage[colorlinks]{hyperref}
\hypersetup{%
        plainpages=true,
        breaklinks=true,
        hypertexnames=false,
        pageanchor=true,
        colorlinks=true,
        linkcolor={blue},
        citecolor={magenta},
        urlcolor={blue},
        anchorcolor={black}
      }
\usepackage[all]{hypcap} 

\newcommand{\figref}[1]{\mbox{Fig.~\ref{#1}}}

\renewcommand{\eqref}[1]{\mbox{Eq.~(\ref{#1})}}

\newcommand{\figurepanel}[2]{Fig.~\hyperref[#1]{\ref*{#1}(#2)}}
\newcommand{\figurepanelNoPrefix}[2]{\hyperref[#1]{\ref*{#1}(#2)}}

\begin{document}

\preprint{APS/123-QED}

\title{Non-Markovian steady states of a driven two-level system}

\author{Andreas Ask}

\affiliation{
 Department of Microtechnology and Nanoscience (MC2), Chalmers University of Technology, SE-41296 Göteborg, Sweden
}%

\author{Göran Johansson}
\affiliation{
 Department of Microtechnology and Nanoscience (MC2), Chalmers University of Technology, SE-41296 Göteborg, Sweden
}

\date{\today}

\begin{abstract}
We show that an open quantum system in a non-Markovian environment can reach steady states that it cannot reach in a Markovian environment. As these steady states are unique for the non-Markovian regime, they could offer a simple way of detecting non-Markovianity, as no information about the system’s transient dynamics is necessary. In particular, we study a driven two-level system (TLS) in a semi-infinite waveguide. Once the waveguide has been traced out, the TLS sees an environment with a distinct memory time. The memory time enters the equations as a time delay that can be varied to compare a Markovian to a non-Markovian environment. We find that some non-Markovian states show exotic behaviors such as population inversion and steady-state coherence beyond $1/\sqrt{8}$, neither of which is possible for a driven TLS in the Markovian regime, where the time delay is neglected. Additionally, we show how the coherence of quantum interference is affected by time delays in a driven system by extracting the effective Purcell-modified decay rate of a TLS in front of a mirror. 
 
\end{abstract}

\maketitle

\section*{\label{sec:level1}}

There are no truly closed quantum systems. In one way or another, a quantum system is always in contact with a noisy environment and will inevitably lose its quantum properties \cite{breuer2007theory}. If the dynamics are Markovian in nature, the environment can be considered memoryless, and there is no back-flow of information. Such systems are described by a quantum dynamical semi-group, whose generator governs a master equation in Lindblad form \cite{Lindblad1976,Gorini1976}. In many realistic systems, the requirements for a Markovian time evolution, such as weak interaction and short environment correlation-times, are not satisfied, and long-time memory effects of the environment influence the system dynamics. In what ways such interaction affects the evolution of a quantum system is not only interesting from a fundamental perspective, but could also prove useful to probe properties of the environment \cite{Cialdi2014,Gessner2014}, and ultimately lead to a better understanding of the decoherence-mechanisms of quantum systems \cite{Breuer2016}. 

Although one cannot translate the classical definition of a Markov process directly to the quantum regime \cite{Vacchini2011}, several definitions and corresponding measures of non-Markovianity for open quantum systems have been introduced \cite{Wolf2008,Breuer2009,Rivas2010,Hou2011,Luo2012,Vasile2011}. These measures are all constructed to detect deviations from Markovianity by characterizing the system's transient dynamics. Irrespective of definition, it has remained an open question if non-Markovianity can be seen in the steady state of driven systems. That would not only simplify the characterization of non-Markovianity in such cases, but it is also an interesting question in itself.

In this Letter, we show that an open quantum system coupled to a non-Markovian environment can reach a unique set of steady states that are out of reach for the system coupled to a Markovian environment. We call these states ``non-Markovian steady states'', as they can be distinguished from \emph{any} state in the Markovian regime. To quantify these steady states' uniqueness, we propose a measure based on trace distance and distinguishability \cite{Fuchs1999}. We note that our measure does not attempt to quantify the degree of non-Markovianity in these systems but rather gives a quantitative measure on how easily one can distinguish these states from the states in the Markovian regime. 

\begin{figure}[t!]
    \centering
    \includegraphics[width = \columnwidth]{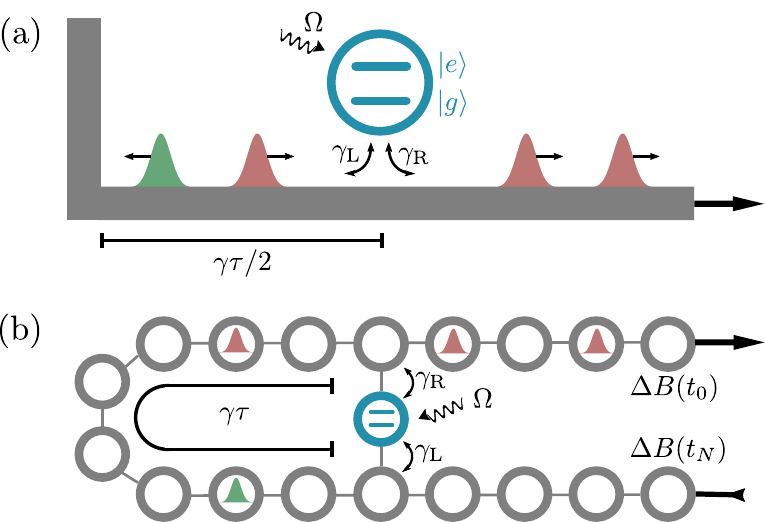}
    \caption{(a) Schematic of an atom in front of a mirror. Photons emitted to the left is reflected from the mirror and interacts with the atom again, giving the atom's environment an effective memory time. (b) Representation of one time step in the evolution of the setup in (a) as a MPS. The waveguide is represented by time bins (gray), which moves together in a convey belt fashion one time step, $\Delta t$, at a time, and interacts with the atom (turquoise) twice.}
    \label{fig:setup}
\end{figure}

To demonstrate when non-Markovian steady states can occur, we study a driven two-level system (TLS) in a semi-infinite waveguide (an atom in front of a mirror) \cite{Dorner2002,Tufarelli2013,Meschede1990,Beige2002,Koshino2012,Wang2012,Pichler2016,Pichler2017,Wiegand2020}, see Fig.~\ref{fig:setup}. The drive amplitude and the system's coupling strength to the waveguide are taken as fixed parameters throughout the system evolution. Once the waveguide has been traced out, the distance to the mirror gives the environment seen by the TLS a distinct memory time. This memory time enters the equations for the system dynamics as a time delay, which can be set to zero for comparison between a Markovian and a non-Markovian environment. Thus, the physical origin of any non-Markovian effects in this system has an easy interpretation in terms of coherent quantum feedback. Additionally, an atom in front of a mirror has been realized with both artificial and natural atoms in a variety of systems already \cite{Eschner2001,Wilson2003,Dubin2007,Hoi2013,Hoi2018,Wen2019,Yong2021}, so the physics discussed in this letter could be further investigated experimentally immediately. We also note that a similar system to the atom in front of a mirror (in fact, they are fully mappable to each other in some parameter regimes) is the giant atom \cite{FriskKockum2014,Guo2017,Kockum2018,Ask2019,Gou2020}, which was recently realized in both the Markovian \cite{Kannan2020,Vadiraj2020} and non-Markovian regime \cite{Andersson2019}.

Despite being an archetypical quantum-optical model-system for decades \cite{Dorner2002}, the atom in front of a mirror has remained a hard system to simulate without resorting to substantial approximations. The propagation time-delay between the atom and the mirror prohibits a treatment based on Markovian master equations, and earlier work on non-Markovianity has been limited to either few excitations \cite{Tufarelli2014,Fang2018} or short time scales \cite{Haikka2010,Haikka2011}. It was only recently that Pichler {\it et al.}~\cite{Pichler2016} proposed a method based on Matrix Product States (MPS) which could allow the system to be integrated all the way to steady state, while still allowing for many excitations in the feedback loop and long delay times. There, it was shown that the time delay does in fact alter the steady state of the atom. However, that does not imply that the state is {\em unique} for the non-Markovian regime, i.e., that the same state cannot be reached using a different drive strength and neglecting the time delay. It could also happen that a non-Markovian environment reduces the purity of the steady state, in which case the effect of the non-Markovian environment could be captured by adding additional pure dephasing to a fully Markovian treatment. In fact, we find that the non-Markovian environment mostly produces steady states one cannot distinguish from those reachable in the Markovian environment. For some system parameters, however, we find non-Markovian steady states that are not only unique for the non-Markovian regime but also show exotic behaviors such as population inversion in the TLS or steady-state coherence beyond $1/\sqrt{8}$, neither of which is possible in the Markovian regime.

\begin{figure}[ht!]

    \includegraphics{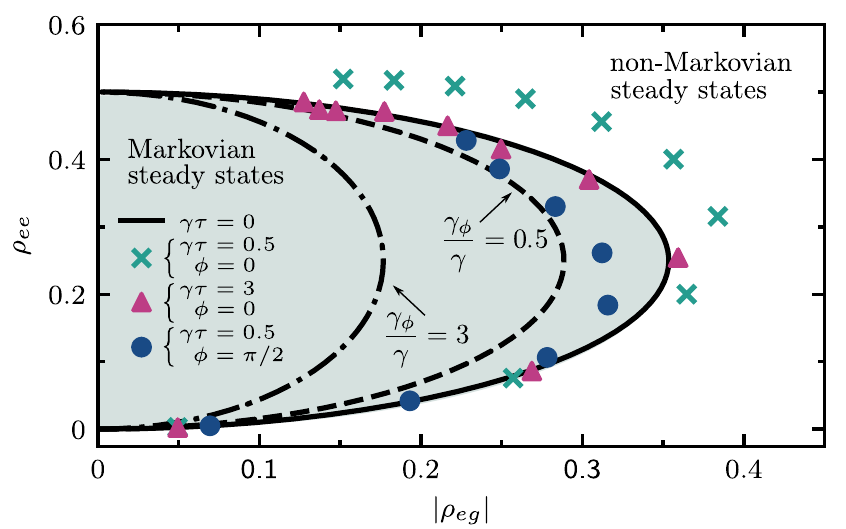}
            \caption{Markovian versus non-Markovian regimes of steady states for a driven TLS in front of a mirror. By neglecting the time delay, the system evolves according to a Markovian master equation and can only reach states lying either on the solid black line (for no pure dephasing) or its inside (with dephasing). If the time delay is taken into account the system can reach steady states which are, e.g., outside of the Markovian regime (green crosses), precisely on the boundary between the two regimes (magenta triangles), or well inside the Markovian regime (blue circles). In all cases the following parameters were used: $\gamma = \gamma_{\text{L}} + \gamma_{\text{R}} = 1$, $\gamma_{\text{L/R}} = \gamma/2$, and $ \Omega/\gamma = [0.1,4]$ ($\Omega/\gamma = [0.1,3.5]$ for $\phi = \pi/2$). 
            }
    \label{fig:markovian_bloch}
\end{figure}

\textit{Definition of non-Markovian steady states}: 
In the Markovian regime, the dynamics of the TLS is given by the Markovian master equation in Lindblad form \cite{Lindblad1976},
\begin{equation}
\begin{split}
    \dot{\rho}_M &= - i[H_{\text{TLS}},\rho] + \frac{\gamma}{2} \left ( 2\sigma^-\rho \sigma^+ - \{\sigma^+\sigma^-,\rho\} \right ) \\
     & + \gamma_{\phi} \left ( 2\sigma^+\sigma^- \rho \sigma^+\sigma^- - \{\sigma^+\sigma^-,\rho\} \right ),
     \label{eq:master}
\end{split}
\end{equation}
where $\gamma_{\phi}$ is a pure dephasing rate, $\gamma$ is a renormalized decay rate due to the mirror $\gamma = 2\gamma'\cos(\phi)$, where $\gamma'$ is the bare decay rate (without the mirror), $\phi$ is a phase shift, $H_{\text{TLS}} = \Delta\sigma^+\sigma^- + \frac{\Omega}{2}(\sigma^+ + \sigma^-)$, where $\Delta = \omega_d - \omega_0$ is the detuning between the TLS transition frequency $\omega_0$ and the drive frequency $\omega_d$, $\Omega$ is the amplitude of the driving field, and $\sigma^+ (\sigma^-)$ creates (annihilates) an excitation in the TLS. We always consider resonant driving throughout the paper, $\Delta = 0$. The solutions to \eqref{eq:master} yields an elliptical area in the Bloch sphere of possible steady states, whose outer boundary is given by $\Omega/\gamma = [0,\infty]$ and $\gamma_{\phi} = 0$, see Fig.~\ref{fig:markovian_bloch}. The mirror is thus irrelevant for determining possible steady states; it only re-scales the decay rate. We let $\rho_M(\gamma,\gamma_{\phi})$ denote the steady state solution to \eqref{eq:master} for a fixed drive strength, and $\rho$ denote the steady-state reduced density matrix of the TLS in a non-Markovian environment. Then, the ability to distinguish the non-Markovian regime from the Markovian regime can be captured by
\begin{equation}
    \mathcal{N}_{ss} = \min_{\gamma,\gamma_{\phi}} T\left[\rho,\rho_M(\gamma,\gamma_{\phi})\right],
    \label{eq:measure}
\end{equation}
where $T[\rho,\rho_M] = \frac{1}{2}\Tr\sqrt{\left(\rho - \rho_M\right)^{\dagger}\left(\rho - \rho_M\right)}$ is the trace distance between $\rho$ and $\rho_M$. As the trace distance is closely related to the \emph{distinguishability} of quantum states \cite{Fuchs1999,nielsen2010}, we define a steady state as non-Markovian if $\mathcal{N}_{ss} > 0$. With this definition, a non-Markovian steady state is a state that is unique for the system in a non-Markovian environment. For the atom in front of a mirror, we show that most non-Markovian steady states would correctly be classified as belonging to a non-Markovian system according to the definition of non-Markovianity in Ref. \cite{Breuer2009}. However, we note that a large degree of non-Markovianity does not necessarily correspond to a large $\mathcal{N}_{ss}$.

\textit{Model:} 
To model a non-Markovian environment, we put the driven TLS in a semi-infinite waveguide, see \figurepanel{fig:setup}{a}. After the waveguide has been traced out, the distance to the mirror gives the environment of the TLS a distinct memory time, $\tau$. Emission towards the mirror enters a coherent feedback loop, in which it pick ups a propagation phase, $\phi$, that in our calculation includes any extra phase shift imposed by the field's boundary condition at the mirror. In a frame rotating with the drive frequency, $\omega_d$, the total Hamiltonian has two parts in the interaction picture $H(t) = H_{\text{TLS}} + H_{\text{int}}(t)$, where the interaction Hamiltonian
\begin{equation}
    H_{\text{int}}(t)  = i\left(\sqrt{\gamma_{\text{L}}} b_L^{\dagger}(t) + \sqrt{\gamma_{\text{R}}} b_R^{\dagger}(t)\right) \sigma^- + \text{H.c.},
\end{equation}
can be rewritten in terms of a single bath operator, $b(t)$, since the mirror couples left and right-going modes $b_R(t) = b_L(t-\tau)e^{i\phi}$, 
\begin{equation}
    H_{\text{int}}(t)  = i\left(\sqrt{\gamma_{\text{L}}} b^{\dagger}(t) + \sqrt{\gamma_{\text{R}}} b^{\dagger}(t-\tau)e^{i\phi} \right) \sigma^- + \text{H.c.},
    \label{eq:interaction}
\end{equation}
where $\gamma_{\text{L}}$ and $\gamma_{\text{R}}$ denotes the decay rate into left (L) and right (R)-going modes in the waveguide, respectively, $\phi$ is the phase shift acquired by a photon (or phonon) traveling to the mirror and back. The phase shift is in fact related to the drive-frequency, $\phi = \omega_d \tau$, but we keep it as an independent variable in order to study the effect of the phase shift and delay time separately. The bosonic operator, $b(t)$, obeys the quantum white-noise commutation-relation $[b(t),b^{\dagger}(t')] = \delta(t-t')$, and is defined as $b(t) = \int_B d\omega b(\omega) \text{exp}[-i(\omega - \omega_d)t]$, where $b^{\dagger}(\omega)$ [$b(\omega)$] creates [annihilates] a photon at frequency $\omega$, satisfying the commutation relation $[b(\omega),b
^{\dagger}(\omega')] = \delta(\omega - \omega')$. A full derivation of the interaction Hamiltonian in \eqref{eq:interaction} can be found in Ref. \cite{Pichler2016}. The interpretation, however, is clear: the TLS interacts with a bosonic bath at time $t$, after some time $\tau$ the TLS interacts with the bath again, the state of the bath at this later time has to be the state of the bath at an earlier time $t-\tau$, taking into account the traveling phase acquired during this time. 

The system dynamics is calculated by solving the quantum stochastic Shrödinger equation (QSSE)
\begin{equation}
    i\frac{d}{dt}\ket{\Psi(t)} = H(t)\ket{\Psi(t)},
    \label{eq:QSSE}
\end{equation}
using the MPS method formulated in Ref.~\cite{Pichler2016}. Matrix product states have shown to be efficient representations of 1-D many-body systems \cite{Schollwock2011,Vidal2003,Vidal2004,faithful,Hastings2007,White1992,Stlund1995}. Solving the QSSE is turned into a many-body problem by discretizing time, $t_n = n \Delta t$, turning \eqref{eq:QSSE} into a dynamical map $\ket{\Psi(t_{n+1})} = U_n \ket{\Psi(t_n)}$. Throughout the paper we use a time step much smaller than all other time scales involved: $\Delta t \ll \{1/\gamma, 1/\Omega\}$. The state of the field in the waveguide is represented by time bins, with associated bosonic noise increments $\Delta B(t_n) = \int_{t_n}^{t_{n+1}} b(t)dt$, which fulfills the commutation relation $[ \Delta B(t_n),\Delta B^{\dagger}(t_{n'})] = \Delta t \delta_{n,n'}$. The operator $\Delta B^{\dagger}(t_n)$ can thus be seen as a creation operator for time bin $n$, with a corresponding Fock state defined as: $\ket{n}_l \equiv \frac{ (\Delta B^{\dagger}_l)^n}{\sqrt{n!\Delta t^n}}\ket{\text{vac}}_l$. The unitary, $U_n$, is written in this time-bin formulation as
\begin{equation}
\begin{split}
    U_n &= \text{exp} \left[ i H_{\text{TLS}}\Delta t + (\sqrt{\gamma_{\text{L}}} \Delta B^{\dagger}(t_n)\sigma^- \right. \\
    & + \sqrt{\gamma_R}\Delta B^{\dagger}(t_{n-k})e^{i\phi}\sigma^-   - \text{H.c.}) \left. \right],
    \end{split}
\end{equation}
here $t_k = k\Delta t $ is the feedback time $\tau$. The total quantum state at time $t_n$, for both the bath and the TLS, is then written as
\begin{equation}
    \ket{\Psi(t_n)} = \sum_{i_{\text{T}},i_1,\ldots,i_N} \Psi_{i_{\text{T}},i_1,\ldots,i_N}(t_n) \ket{i_{\text{T}}} \otimes \ket{i_1} \otimes \ldots \otimes \ket{i_N},
\end{equation}
where $t_N = N \Delta t$ is the total integration time, $i_T$ denotes the state of the TLS, and $i_j$ is the photon number in time-bin $j$. The initial state is written as an MPS ansatz
\begin{equation}
    \Psi_{i_S,i_1,\ldots,i_N}(t_0) = M[S]^{i_T}M[1]^{i_1}\ldots  M[N]^{i_N},
\end{equation}
where $M[j]^{i_j}$ is a matrix of dimension $D_j \times D_{j+1}$. The maximum matrix dimension $D_{\text{max}}$ in the MPS chain is referred to as the bond dimension and sets an upper limit to the amount of entanglement in the system, which in our system depend on the length of the feedback loop \cite{Pichler2016}. We use a bond dimension of $D_{\text{max}} = 32$ for the moderate time delays considered here. The state amplitudes are updated in each time step using standard MPS techniques \cite{Schollwock2011,Shi2006}. 

\begin{figure}[h]
        \centering
    \includegraphics{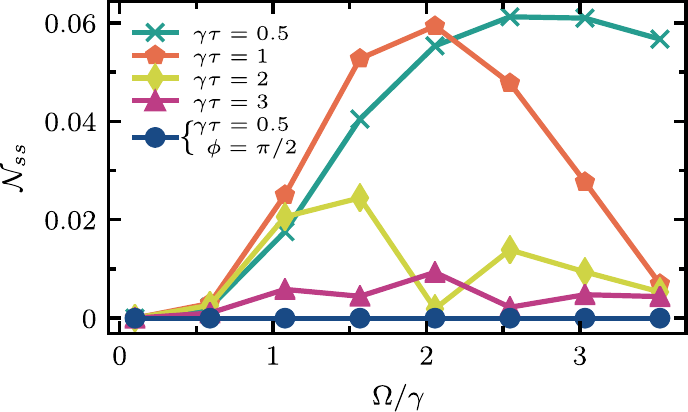}
    \caption{Non-Markovianity of the steady state evaluated using the measure introduced in \eqref{eq:measure}. A phase shift of $\phi=0$ was used unless it is stated otherwise in the figure. A large $\mathcal{N}_{ss}$ means that the state is easily distinguishable from any state in the Markovian regime.}
    \label{fig:measure}
\end{figure}

\textit{Non-Markovian regime:} Two parameters are important for quantifying the significance of the feedback: $\gamma\tau$, where $\gamma = \gamma_{\text{\text{L}}} + \gamma_{\text{R}}$, and $\Omega \tau$. Only when both $\gamma\tau \ll 1$ and $\Omega \tau \ll 1$ does the system not have time to evolve during the feedback time, and a Markovian treatment is valid. The situation is thus different from the non-driven TLS in which the single parameter $\gamma\tau$ determines the non-Markovian properties alone.  

We first study the effect of the time delay, and set $\phi = 0$. The feedback is thus in phase with the drive and they interfere constructively. In all calculations that follows we use $\gamma = 2\gamma_{\text{L}} = 2\gamma_{\text{R}} = 1$. In \figref{fig:markovian_bloch} we plot the steady-state reduced density-matrix elements of the TLS for drive-strengths in the range $\Omega/\gamma = [0.1,4]$ for $\gamma\tau = 0.5$ (green crosses) and $\gamma\tau = 3$ (magenta triangles). The trace distance to the closest Markovian steady state, $\mathcal{N}_{ss}$ [\eqref{eq:measure}], is plotted for a greater variety of time delays in \figref{fig:measure}. From these two figures we make the following observations: (i) When the driving is weak, the steady state cannot be distinguished from a Markovian steady state, independent of time delay. (ii) The states start to deviate from the Markovian regime initially for $\gamma\tau > 0$, reaches a maximum deviation at $\gamma\tau \approx 0.5$, and then starts to approach the Markovian regime again. For sufficiently long time delays the states cannot be distinguished from a Markovian state anymore. (iii) For $\gamma\tau = 0.5$ the TLS shows both population inversion, $\rho_{ee} > 1/2$, for sufficiently strong driving, and larger coherence than what is possible in the Markovian regime, $|\rho_{eq}| > 1/\sqrt{8}$. The oscillatory behavior that can be seen in both \figref{fig:markovian_bloch} and \figref{fig:measure} is due to the additional phase shift, $\Omega\tau$, induced by the drive, primarily noticeable for the longer time delays.

When the phase shift deviates from $0 \mod 2\pi$, the states either approaches or falls well inside the Markovian regime (blue circles in \figref{fig:markovian_bloch}). Inside the Markovian regime $\mathcal{N}_{ss} = 0$ per definition. 

\begin{figure}[ht!]
    \centering
    \includegraphics{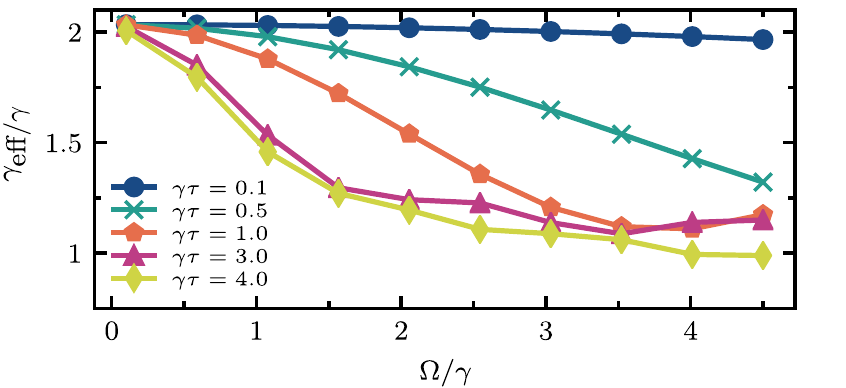}
    \caption{Effective decay rate for $\phi = 0$ as a function of drive strength for various time delays. }
    \label{fig:decay}
\end{figure}

Coherent quantum interference phenomena play an important role in many applications in waveguide QED. It is well known, e.g., that the mirror doubles the TLS' decay rate to the waveguide due to the Purcell effect in the Markovian regime (for $\phi = 0$). Increasing the distance to the mirror reduces the coherence of the radiation that comes back to the TLS. The longer time it takes for the feedback to come back, the higher chance for spontaneous emission to occur in the TLS. For long enough time delay, the TLS should behave as if it was positioned in an infinite waveguide instead, without the mirror present. This is precisely what we observe in \figref{fig:decay}. We extract the ``effective'' decay rate by calculating the ratio between the output time-bin population and the TLS population, $\gamma_{\text{eff}} = \langle \Delta B^{\dagger}_{\text{out}} \Delta B_{\text{out}} \rangle_{ss}/\rho_{ee}$. For weak driving we observe that the decay rate is not affected by the time delay, as the drive strength increases, however, it approaches the expected value of $\gamma$. We note that the effective decay rate could have been extracted from the master equation in \eqref{eq:master} if negative dephasing rates were allowed. In fact, all the non-Markovian steady states seen in \figref{fig:markovian_bloch} could be described by the master equation with a negative dephasing rate. Since negative dephasing rates have been used to describe temporary increases in quantum coherence for non-Markovian systems \cite{Shrikant2018,Cai2020}, we find it interesting to note that such effects can persist all the way to the steady state.

\begin{figure}[t]
    \centering
    \includegraphics{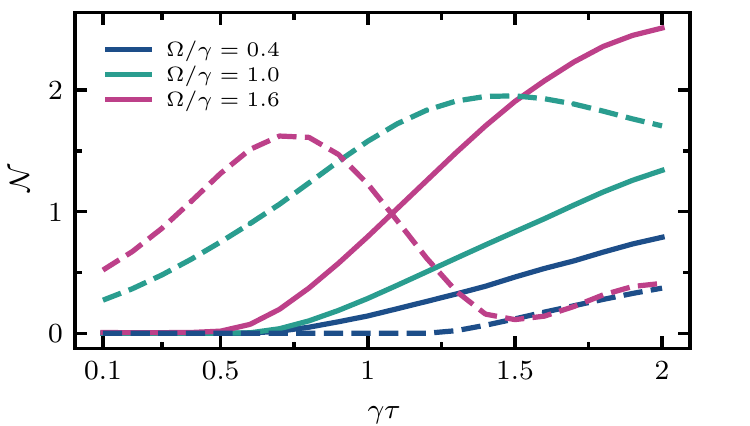}
    \caption{Non-Markovianity according to the measure in \eqref{eq:nonM} as a function of time delay. Dashed lines are for the same drive strengths as written in the figure but for $\phi = \pi/2$. By comparing with \figref{fig:markovian_bloch} we conclude that a large non-Markovianity does not correspond to a steady state that can be distinguished from the Markovian regime.}
    \label{fig:nonM}
\end{figure}

\textit{Non-Markovianity measure of the transient dynamics:} 
Finally, we make the remark that maximizing $\mathcal{N}_{ss}$ is not necessarily the same thing as maximizing the non-Markovianity of the system in a traditional sense. For this argument, we compare $\mathcal{N}_{ss}$ to the measure proposed in Ref.~\cite{Breuer2009}, 
\begin{equation}
    \mathcal{N} = \text{max}_{\rho_{1,2}} \int_{\sigma > 0} dt \sigma(t,\rho_{1,2}),
    \label{eq:nonM}
\end{equation}
where $\sigma(t,\rho_{1,2}) = \frac{d}{dt}T[\rho_1(t),\rho_2(t)]$, and $T[\rho_1,\rho_2]$ denotes the trace distance. The maximum is taken over all pairs of initial states. For our system, we can safely choose the ground and excited states as the two initial states \cite{Wissmann2012}. We plot $\mathcal{N}$ as a function of time delay for various drive-strengths in \figref{fig:nonM} for both $\phi = 0 $ (solid lines) and $\phi = \pi/2$ (dashed lines). In \figref{fig:measure}, we saw the largest $\mathcal{N}_{ss}$ for $\gamma\tau \approx 0.5-1$, which is barely non-Markovian according to $\mathcal{N}$, whereas longer time delays and stronger driving increases $\mathcal{N}$ significantly.

\textit{Conclusion:} 
We have introduced the concept of  ``\emph{non-Markovian steady states}" as a set of steady states unique for open quantum systems in a non-Markovian environment. The system cannot reach these states if it is coupled to a Markovian environment and could thus offer a simple way of detecting non-Markovianity as only a steady state measurement is required. Moreover, we introduced an appropriate measure for these states' uniqueness based on the trace distance to the closest Markovian steady state. As an example, we show that non-Markovian steady states occur in a driven TLS in a semi-infinite waveguide. Among the non-Markovian steady states, we find states with population inversion in the TLS, or steady state coherence larger than $1/\sqrt{8}$, two impossible scenarios in the Markovian regime. Additionally, we showed that time delay could have a detrimental effect on coherent quantum interference in waveguides by extracting the effective Purcell-modified decay-rate as a function of time delay and drive strength. 

\textit{Acknowledgments:} We thank Arne L. Grimsmo for valuable discussions, and acknowledge funding from the Swedish Research Council (VR) through Grant No.~2017-04197. G.J. also acknowledges funding from the Knut and Alice Wallenberg foundation (KAW) through the Wallenberg Centre for Quantum Technology (WACQT).


\bibliography{apssamp}

\end{document}